\documentstyle[prl,aps,epsfig,multicol]{revtex}

\def\Journal#1#2#3#4{{#1}{\bf #2}, #3 (#4)}

\def\NIMA{{Nucl. Instrum. Methods in Phys. Research}~{\bf A}}
\def\NPA{{Nucl. Phys.}~{\bf A}}

\def\PLB{{Phys. Lett.}~{\bf B}}

\def\PRL{Phys. Rev. Lett.\ }

\def\PRC{{Phys. Rev.}~{\bf C}}

\begin{document}
\draft


\title{Elliptic Flow of Identified Hadrons in Au+Au
Collisions at $\sqrt{s_{_{\rm NN}}}$ = 200~GeV}


\author{
S.S.~Adler,$^{5}$
S.~Afanasiev,$^{17}$
C.~Aidala,$^{5}$
N.N.~Ajitanand,$^{43}$
Y.~Akiba,$^{20,38}$
J.~Alexander,$^{43}$
R.~Amirikas,$^{12}$
L.~Aphecetche,$^{45}$
S.H.~Aronson,$^{5}$
R.~Averbeck,$^{44}$
% ...
\\(PHENIX Collaboration)\\
}
\address{
% ...
$^{5}$Brookhaven National Laboratory, Upton, NY 11973-5000, USA\\
% ...
$^{12}$Florida State University, Tallahassee, FL 32306, USA\\
% ...
$^{17}$Joint Institute for Nuclear Research, 141980 Dubna, Moscow Region, Russia\\
% ...
$^{20}$KEK, High Energy Accelerator Research Organization, Tsukuba-shi, Ibaraki-ken 305-0801, Japan\\
% ...
$^{38}$RIKEN (The Institute of Physical and Chemical Research), Wako, Saitama 351-0198, JAPAN\\
% ...
$^{43}$Chemistry Department, Stony Brook University, SUNY, Stony Brook, NY 11794-3400, USA\\
$^{44}$Department of Physics and Astronomy, Stony Brook University, SUNY, Stony Brook, NY 11794, USA\\
$^{45}$SUBATECH (Ecole des Mines de Nantes, CNRS-IN2P3, Universit{\'e} de Nantes) BP 20722 - 44307, Nantes, France\\
% ...
}

\date{\today}
\maketitle

\begin{abstract}

The anisotropy parameter ($v_2$), the second harmonic of the
azimuthal particles distribution, has been measured with the PHENIX
detector in Au+Au collisions at $\sqrt{s_{_{NN}}} = $ 200~GeV for
identified and inclusive charged particle production at central rapidities
($|\eta| < 0.35$) with respect to the reaction plane defined at high
rapidities ($|\eta| = \; $3--4).  We observe that the $v_2$ of
mesons falls below that of (anti)baryons for $p_{\rm T} >$ 2~GeV/c,
in marked contrast to the predictions of a hydrodynamical model.
A quark coalescence model is also investigated.

\end{abstract}



\begin{multicols}{2}   
\narrowtext            

Event anisotropy is expected to be sensitive to the early stage of
ultra-relativistic nuclear collisions at the Relativistic Heavy Ion
Collider (RHIC).  The possible formation of a quark-gluon plasma (QGP)
could affect how the initial anisotropy in coordinate space is
transferred into momentum space in the final state.  
The anisotropy parameter $v_2$ for a selection of produced particles
is derived from the azimuthal distribution of those particles.

\begin{equation}
\frac{dN}{d\phi} \; \propto \; 1 \; + \; 2 \, v_{2} \,
\cos 2 ( \phi - \Phi_{\rm RP})
\end{equation}

\noindent
where $\phi$ is the azimuthal direction of the particle and 
$\Phi_{\rm RP}$ is the
direction of the nuclear impact parameter (``reaction plane'') in a
given collision.  Measurements of the parameter $v_2$ in RHIC
collisions have been performed
\cite{star01,star02,star03,star04,star05,phnx01} for charged particles
and for identified particles.  The current work reports results for
charged particles versus transverse momentum ($p_{\rm T}$) out to 5~GeV/c, 
and extends previous measurements for identified particles
out to 3~GeV/c for $\pi$ and K, and to 4~GeV/c for protons.
(Previous measurements of the $v_2$ for $\pi$, K, and p
extended to 1~GeV/c at $\sqrt{s_{_{NN}}} = $ 130~GeV.\cite{star02})
Detailed measurements of the azimuthal anisotropy are important to 
eventually discriminate among different possible
scenarios for its physical origin.  Such scenarios include:
hydrodynamical flow of compressed hadronic matter, the production of
multiple mini-jets, and an anisotropy developed during an early
quark-matter phase of the collision.  It has been observed that $v_2$
saturates at $p_{\rm T}$ $\sim$ 2~GeV/c and above
\cite{star04,star05}.  The cause of this saturation is not yet known;
however, we note that at this momentum the particle composition
is very different than at low momentum in that the proton yield is
comparable to the pion yield \cite{phnx02}.  This makes the
measurement of $v_2$ for separately identified particles especially
interesting.

The measurements described here were carried out in the PHENIX
experiment at RHIC \cite{phnx08}.  About 28~M minimum bias Au+Au
collisions at $\sqrt{s_{_{\rm NN}}}$ = 200~GeV from the 2001-2002 run period
 (Run-2) are used
in the analysis. Charged particles are measured in the
central arm spectrometers ($|\eta| < 0.35$) \cite{phnx13} where PHENIX
 has excellent particle identification capabilities\cite{phnx03}. The drift
chamber (DC) and the first pad chamber plane (PC1) together with the
collision vertex define the charged particle tracks. 
%
%
In order to reduce background, the reconstructed tracks are confirmed by
requiring matching hits in the outer detectors, {\em i.e.} the third
pad chamber plane (PC3) and the electromagnetic calorimeter (EMCAL) or
the time-of-flight detector (TOF).  In this analysis, the TOF detector is 
used to identify charged particles up to 4~GeV/c in $p_{\rm T}$. 
Particle time-of-flight is measured using the TOF with respect
to the collision time defined by beam counters (BBC), and is used to
calculate mass squared using the particle momentum and the flight path
length \cite{phnx02}. The timing resolution of the system is $\simeq$
120 ps.  A momentum dependent $\pm 2$ $\sigma$ cut on mass squared
allows particle identification in the $p_{\rm T}$ range 0.2 $<$
$p_{\rm T}$ $<$ 3~GeV/c for pions, 0.3 $<$ $p_{\rm T}$ $<$ 3~GeV/c for
kaons, and 0.5 $<$ $p_{\rm T}$ $<$ 4~GeV/c for protons.  The
contamination of mis-identified particles is less than 10$\%$.  In
addition to collision time, the BBC provide z-vertex position
information.  
The two beam counters are located at $|z|$=1.5~m from
the collision point, covering $|\eta| = 3 \sim 4$.
They consist of 64 photo-multiplier tubes (PMT) equipped with quartz 
Cherenkov radiators in front surrounding the beam pipe.
The large charged
multiplicity (a few hundred) in $|\eta| = 3 \sim 4$ and the non-zero
signal of event anisotropy in this $\eta$ range enables us to estimate
the azimuthal angle of the reaction plane in each event using the BBC
with full azimuthal angle coverage.

\begin{figure}
\centerline{\epsfig{file=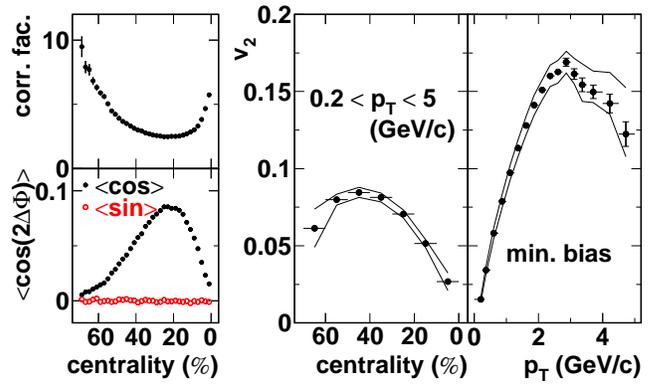,width=1.0\linewidth}}
\caption[]{(color online) 
Correlation of reaction planes between two beam counters
for the second moment is shown as a function of centrality (bottom-left) 
and the correction factor for the combined reaction plane resolution 
of two beam counters is shown as a function of centrality (top-left). 
The value of $v_2$ for charged particles
is shown as a function of centrality (middle) 
and as a function of $p_{\rm T}$ (right).}
\label{prl_fig1}
\end{figure}

Since the $v_2$ parameter is in effect a quadrupole moment, the
anisotropy which gives rise to a non-zero $v_2$ is often referred to
as an ``elliptic flow.''  It is extracted by first determining the
reaction plane angle $\Phi_{\rm RP}$ for each event,

\begin{equation}
\tan 2 \Phi_{\rm RP} = \frac{\Sigma n_{\rm ch} \sin 2 \phi_{\rm PMT}}
                            {\Sigma n_{\rm ch} \cos 2 \phi_{\rm PMT}}
\end{equation}

\noindent
where $n_{\rm ch}$ is the number of charged particles per PMT
(determined from the pulse height in each PMT) and 
$\phi_{\rm PMT}$ is the azimuthal angle of each PMT. Then, it is calculated
by the Fourier moment $v_{2} = \langle \cos~ 2(\phi - \Phi_{\rm RP}) \rangle$
over all particles, for all events in a given sample \cite{volos1}.
Corrections \cite{volos1,e87701,e87702,e87703}
are applied to account for finite resolution
in the reaction plane determination, and for possible azimuthal 
asymmetries in the reaction plane detector response.
The bottom-left panel in Fig.~\ref{prl_fig1} shows the average 
cosine of the difference between the two reaction planes defined 
by the beam counters at $\eta$ = $3 \sim 4$ and at $\eta$ = 
$-4 \sim -3$ using the the elliptic (second) moment definition. 
In order to improve the reaction plane resolution, a 
combined reaction plane is defined by averaging the reaction 
plane angles obtained from each BBC, using the elliptic moment
in each.
The estimated resolution of the combined reaction plane\cite{volos1},  
$\langle \cos~ 2(\Phi_{\rm measured}-\Phi_{\rm true}) \rangle$,
%
%
has an average of 0.3 over centrality with a maximum of about 0.4. 
The estimated correction factor, which is the inverse of the 
resolution for the combined reaction plane, is shown
in the top-left panel in Fig.~\ref{prl_fig1}.

The present technique is distinguished by defining the reaction
plane angle using particles at high rapidity when measuring 
$v_2$ for particles at mid-rapidity.  Other measurements
of $v_2$ for mid-rapidity particles at RHIC have used reaction
planes defined with mid-rapidity particles; or have employed a
technique of measuring angular correlations between pairs of 
particles at mid-rapidity.  While these different approaches
generally seek to measure the same thing, they are not identical
and a variety of physics effects can cause them to yield
different results from the same collision sample 
\cite{olli01,olli02,star04}. 
Because of the large rapidity gap between the reaction plane 
and the mid-rapidity acceptance of about 3 units, it is expected 
that this analysis is less affected by non-flow contributions. 
However, we do not observe any
substantial difference between the $v_2$ results shown here 
and published results for the $v_2$ of
charged particles at RHIC in the $p_{\rm T}$ range
where they are available. 

The centrality of each collision is defined using the simultaneous
measurement of the total number of particles measured in the BBC
and the total energy measured in the zero degree calorimeter \cite{phnx04}. 
The middle panel in Fig.~\ref{prl_fig1} shows the centrality 
dependence of $v_2$ for charged particles measured at 
mid-rapidity ($|\eta| < 0.35$) with respect to the reaction plane
defined above. The centrality is measured in percentile from the 
most central collision. The $v_2$ parameter decreases for both 
peripheral and central collisions with a maximum at about 50$\%$ 
of the geometric cross section. Beyond 70$\%$, the correction 
factor due to the reaction plane resolution 
is large, as shown the left-most panel in the Fig.~\ref{prl_fig1}.  
This limits the centrality range used in this analysis.  

The right-most panel in Fig.~\ref{prl_fig1} shows 
the transverse momentum dependence of $v_2$ for charged 
particles with respect to the reaction plane for minimum-bias events. 
The data above a $p_{\rm T}$ of 2~GeV/c clearly show a deviation from 
the monotonically increasing behavior seen at smaller $p_{\rm T}$. 
The systematic
errors are shown as line bands, which are estimated by several 
reaction plane methods using the two single beam counters or combined
beam counters and by several different ways to correct non-uniform
reaction plane distribution: ``inverse weighting,'' ``re-centering of
sine and cosine summation,'' ``Fourier expansion'' and combinations of
those above \cite{volos1,phnx05}. 
Those systematic errors are estimated to be about 10$\%$, depending on 
centrality, and are independent of $p_{\rm T}$.  Above 3~GeV/c,
background tracks result in an additional systematic error of
about 10$\%$, depending on $p_{\rm T}$, which is included
in the upper error band \cite{phnx15}.

\begin{figure}
\centerline{\epsfig{file=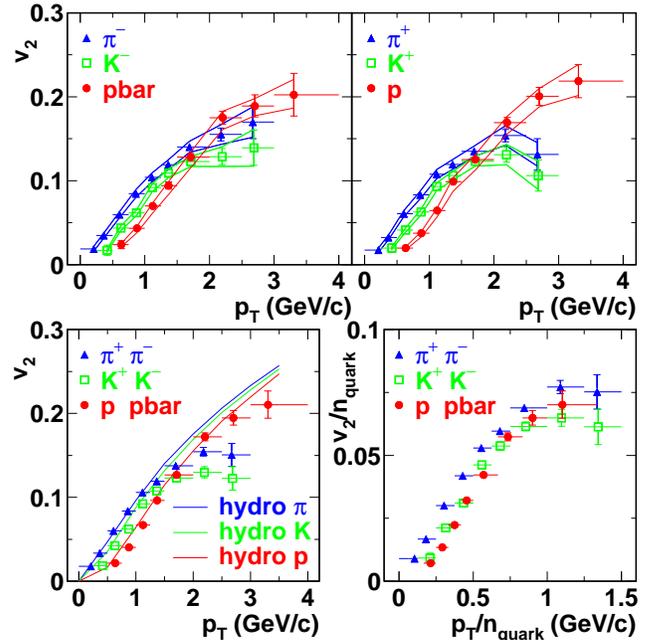,width=1.0\linewidth}}
\caption[]{(color online) Transverse momentum dependence of $v_2$ for 
identified particles, $\pi^-$, $K^-$, $\overline{p}$ (top-left) and $\pi^+$, 
$K^+$, $p$ (top-right). The circles show $p$ and $\overline{p}$, 
the squares show $K^+$ and $K^-$, and the triangles show $\pi^+$ 
and $\pi^-$ for minimum bias events. 
Statistical errors are represented by 
error bars and overall systematic error due to all sources by the solid lines 
in the top two panels.  The combined positive 
particles and negative particles are shown in the bottom-left panel,
and the lines there represent the result of a hydrodynamical 
calculation \cite{houv01} including a first-order phase transition 
with a freeze out temperature of 120 MeV for $\pi$, $K$ and $p$
from upper to lower curves, respectively.
The bottom-right panel shows the quark $v_2$ as a function of 
the quark $p_{\rm T}$ by scaling both axes with the number
of quarks for each particle, as motivated by a quark 
coalescence model \cite{volos2}.}
\label{prl_fig2}
\end{figure}

In Fig.~\ref{prl_fig2}, the transverse momentum dependence of $v_2$
for identified particles is shown.  The top-left panel shows
negatively charged particles, while the top-right panel shows
positively charged particles as described in the figure caption. The
statistical errors and the systematic errors are plotted
independently.  From the lambda particle spectra measured in the
PHENIX central arm, it is determined that approximately 35$\%$ of the
protons originate from lambda decays (``lambda
feed-down'')\cite{phnx06}. The effect of the lambda feed-down on the
measured $v_2$ of the proton is studied by varying the lambda $v_2$
with Monte Carlo simulation.  Protons resulting from lambda feed-down
increase the measured $v_2$ value.  Using the value of the lambda
$v_2$ measured at $\sqrt{s_{_{\rm NN}}}$ = 130~GeV at RHIC \cite{star03},
the effect on the
proton $v_2$ would be less than 10$\%$. Less than 5$\%$ of protons
originate from decays of particles not involving the lambda.  Based on
further simulations of their decays to protons, we estimate that the total
systematic error due to feed-down is at most 11$\%$ depending on $p_{\rm T}$,
which is included in the lower sytematic error band in Fig.~\ref{prl_fig2}.

The combined positive and negative particles are shown in the
bottom-left panel.  The lines in that panel represent a
hydrodynamical calculation\cite{houv01} 
including a first-order phase transition
with a freeze-out temperature of 120 MeV. The data show that at lower
$p_{\rm T}$ ($< 2$ GeV/c), the lighter mass particles have a larger
$v_2$ at a given $p_{\rm T}$, which is reproduced by the model
calculations.  We note, however, that the difference between the charged
kaons and charged pions is larger than the model predicts.  

A striking feature observed at higher $p_{\rm T}$ is that
the $v_2$ of $p$ and $\overline{p}$ are larger than for $\pi$ and $K$
at $p_{\rm T}$ $>$ 2~GeV/c.  This is in sharp contrast to the hydrodynamical
picture, which would predict the same mass-ordering for $v_2$ at
all $p_{\rm T}$.  In our data the mesons begin to show a departure
from the hydrodynamical prediction at $p_{\rm T}$ of about 1.5~GeV/c,
while the (anti)baryons agree with the prediction up until 3~GeV/c
but may be deviating at higher $p_{\rm T}$.
Such behavior is predicted by the quark coalescence mechanism
\cite{volos2}, as shown in the bottom-right panel where both $v_2$
and $p_{\rm T}$ have been scaled by the number of quarks. This could be an 
indication that the $v_2$ of measured hadrons is already established in a
quark-matter phase, although it does not explain why the quark $v_2$
would saturate with $p_{\rm T}$. There exist other scenarios that could be
applicable at RHIC, but we have selected two simple models (hydrodynamical and 
quark coalesence) only to emphasize the experimental evidence of the
crossing of $v_2$ for mesons and baryons.

\begin{figure}
\centerline{\epsfig{file=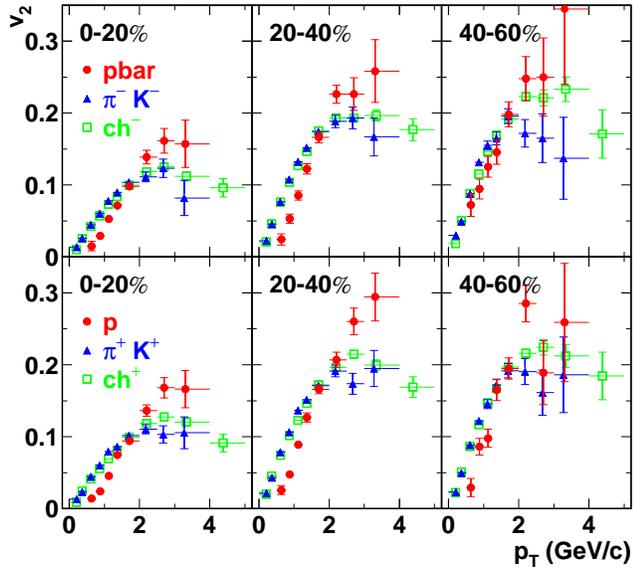,width=1.0\linewidth}}
\caption[]{(color online) Transverse momentum dependence of $v_2$ for 
combined $\pi^-$ and $K^-$ (top) or $\pi^+$ and $K^+$ (bottom) 
compared with $\overline{p}$ (top) and $p$ (bottom).
In addition, results for inclusive negative (top)
and positive (bottom) charged particle distributions 
are plotted as open squares.
From the left to right, the different centrality selections 
are shown for 0-20$\%$ (left), 20-40$\%$ (middle) and 
40-60$\%$ (right).}
\label{prl_fig3}
\end{figure}

As an additional illustration of the different behavior for mesons and
baryons, the transverse momentum dependences of the $v_2$ parameter
are shown in Fig.~\ref{prl_fig3} for different particles and
different centralities.  Since the particle identification separation
of $K$ and $p$ goes up to 4~GeV/c, the combined $\pi$ and $K$ can be
compared with protons up to 4~GeV/c.  The charged particle acceptance
is larger than the TOF acceptance where the particle identification
can be performed.  Therefore, the statistical fluctuations for the
charged particle $v_2$ are smaller than for the $p$, $\overline{p}$
and $\pi+K$.  The trend exhibited in Fig.~\ref{prl_fig2} for minimum
bias spectra, in which the $v_2$ parameter for (anti)baryons exceed
those for mesons at $p_{\rm T}$ $>$ 2~GeV/c, is shown here to occur
for all centralities.


In summary, the value of the $v_2$ parameter for identified and
inclusive charged particle production at mid-rapidity has been measured 
with respect to the reaction plane defined in the forward and backward
rapidity regions in $\sqrt{s_{_{\rm NN}}}$ = 200~GeV Au+Au collisions, using the PHENIX
experiment at RHIC.  The value of $v_2$ for charged
particles decreases for both peripheral and central collisions with a
maximum at about the 50$^{\rm th}$ percentile of the geometric cross
section.  We have observed that for charged particles $v_2$
increases with $p_{\rm T}$ up to about 2~GeV/c, then starts to
saturate or decrease slightly.
However, the detailed behavior is different for different
particle species. The lighter particles have larger $v_2$ than the
heavier particles for $p_{\rm T}$ below 2~GeV/c.  This trend is partly
reversed above 2~GeV/c where the proton and anti-proton have larger
$v_2$ than mesons, a pattern which persists over all centralities.  A
hydrodynamical calculation can reproduce the mass-ordering and
magnitude of $v_2$ for the different particles in the region up to
2~GeV/c, but fails to reproduce either in the $p_{\rm T}$ region above
2~GeV/c.  As an alternative, we investigated the quark-coalescence
scenario, in which the anisotropy of the final-state hadrons is largely
inherited from the anisotropy of quarks in a preceding quark-matter
phase.  The quark-coalescence model makes a definite prediction for a
simple scaling behavior between the $v_2$ for mesons and for
(anti)baryons, and this scaling behavior is largely, though not
perfectly, borne out in our data.  
%
%
Further measurements extending to
higher $p_{\rm T}$ involving more identified species will be required
to discriminate among alternative scenarios for the origin of elliptic
flow at RHIC.


We thank the staff of the Collider-Accelerator and Physics
Departments at BNL for their vital contributions.  We acknowledge
support from the Department of Energy and NSF (U.S.A.), MEXT and
JSPS (Japan), CNPq and FAPESP (Brazil), NSFC (China), CNRS-IN2P3
and CEA (France), BMBF, DAAD, and AvH (Germany), OTKA (Hungary), 
DAE and DST (India), ISF (Israel), KRF and CHEP (Korea),
RAS, RMAE, and RMS (Russia), VR and KAW (Sweden), U.S. CRDF 
for the FSU, US-Hungarian NSF-OTKA-MTA, and US-Israel BSF.








\end{multicols}    


\begin{references}
\bibitem[\ast]{Deceased}Deceased     
\bibitem[\dagger]{Spokesperson}Spokesperson: zajc@nevis.columbia.edu
\bibitem{star01} K.H. Ackermann et al.,
                 \Journal{\PRL} {86}{402}{2001}.
\bibitem{star02} C. Adler et al.,
                 \Journal{\PRL} {87}{182301}{2001}.
\bibitem{star03} C. Adler et al.,
                 \Journal{\PRL} {89}{132301}{2002}.
\bibitem{star04} C. Adler et al.,
                 \Journal{\PRC} {66}{034904}{2002}.
\bibitem{star05} C. Adler et al.,
                 \Journal{\PRL} {90}{032301}{2003}.
\bibitem{phnx01} K. Adcox et al., 
                 \Journal{\PRL} {89}{212301}{2002}.
\bibitem{phnx02} K. Adcox et al., 
                 \Journal{\PRL} {88}{242301}{2002}.
\bibitem{phnx08} K. Adcox et al., 
                 \Journal{\NIMA} {499}{469}{2003}.
\bibitem{phnx13} J. T. Mitchell et al., 
                 \Journal{\NIMA} {482}{491}{2002}.
\bibitem{phnx03} H. Hamagaki et al., 
                 \Journal{\NPA} {698}{412}{2002}.
\bibitem{volos1} A. Poskanzer and S. Voloshin, 
                 \Journal{\PRC} {58}{1671}{1998}.
\bibitem{e87701} J. Barrette et al., 
                 \Journal{\PRL} {73}{2532}{1994}.
\bibitem{e87702} J. Barrette et al., 
                 \Journal{\PRC} {55}{1420}{1997}.
\bibitem{e87703} J. Barrette et al., 
                 \Journal{\PRC} {56}{3254}{1997}.
\bibitem{olli01} N. Borghini, P.M. Dinh and J.Y. Ollitrault, 
                 \Journal{\PRC} {63}{054906}{2001}.
\bibitem{olli02} N. Borghini, P.M. Dinh and J.Y. Ollitrault, 
                 \Journal{\PRC} {64}{054901}{2001}.
\bibitem{phnx04} K. Adcox et al., 
                 \Journal{\PRL} {86}{3500}{2001}.
\bibitem{phnx05} S. Esumi et al.,
                 \Journal{\NPA} {715}{599}{2003}.
\bibitem{phnx15} K. Adcox et al., 
                 \Journal{\PRL} {88}{022301}{2002}.
\bibitem{phnx06} K. Adcox et al., 
                 \Journal{\PRL} {89}{092302}{2002}.
\bibitem{houv01} P. Huovinen, P.F. Kolb, U.W. Heinz, P.V. Ruuskanen and 
                 S.A. Voloshin,  
                 \Journal{\PLB} {503}{58}{2001}.
\bibitem{volos2} D. Molnar and S. Voloshin, nucl-th/0302014.


\end{references}
\end{document}